\documentstyle[11pt,aaspp,epsf]{article}
\begin{document}
\title{ULTRA-LOW FREQUENCY GRAVITATIONAL RADIATION from MASSIVE BLACK HOLE
BINARIES}
\author{Mohan Rajagopal and Roger W. Romani\altaffilmark{1}}
\affil{Stanford University}

\authoraddr{Physics Dept., Stanford University, Stanford CA  94305-4060}

\altaffiltext{1}{Alfred P. Sloan Fellow}

\begin{abstract}
	For massive black hole binaries produced in galactic mergers,
we examine the possibility of inspiral induced by interaction with
field stars. We model the evolution of such binaries for a range of
galaxy core and binary parameters, using numerical results from the
literature to compute the binary's energy and angular momentum loss
rates due to stellar encounters and including the effect of
back-action on the field stars.  We find that only a small fraction of
binary systems can merge within a Hubble time via unassisted stellar
dynamics.  External perturbations may, however, cause efficient
inspiral. Averaging over a population of central black holes and
galaxy mergers, we compute the expected background of gravitational
radiation with periods $P_w \sim 1-10$y. Comparison with sensitivities
from millisecond pulsar timing suggests that the strongest sources may
be detectable with modest improvements to present experiments.
\end{abstract}

\keywords{black holes --- galaxies:interactions --- pulsars:timing ---
radiation mechanisms:gravitational}

\section{Introduction}

The possibility that many galaxies contain a central black hole of
$10^7 M_\odot$ or more, coupled with the ideas that elliptical
galaxies probably formed from mergers ({\it c.f.}  Barnes \& Hernquist,
1992) and that all galaxies may have formed from multiple mergers of
subunits, has led naturally to consideration of the fate of two or
more massive black holes in a merger remnant.  Simulations suggest
that the cores of victim galaxies merge via violent relaxation in a
few crossing times to form the remnant core (Barnes \& Hernquist);
with some possible exceptions (Governato {\it et al.} 1994), black
holes in the victims then end up in the core of an elliptical or a
spiral bulge.  The holes sink towards the center of the stellar
distribution on the dynamical friction time scale of $\sim 10^6$ years
until they become bound to one another (Begelman, Blandford \& Rees
1980, hereafter BBR), forming a massive black hole binary (BHB).  If
the binary loses enough energy and angular momentum to the field
stars, which is by no means certain, it will enter a regime where
gravitational radiation alone can bring about inspiral
and coalescence within the Hubble time.

	During BHB inspiral, the gravity waves generated sweep through
a range of frequencies. In the final merger of two $10^9M_\odot$
black holes, a `chirp' of gravitational radiation with periods as small
as $P_{GR} \sim 10^4$s will be emitted.  It has been suggested that
this radiation might be detected through the Doppler tracking of
spacecraft and by using space-based interferometers (Thorne \& Braginsky
1976, Thorne 1992, Fukushige {\it et al.} 1992a). Unfortunately, the binary
spends only few wave periods in the final chirp, so if inspiral is rare such
events will be difficult to observe. In contrast, at binary
frequencies $\sim 0.01 - 1\;\;\mu$-Hertz, the BHB persists for many
orbits. As first pointed out by Detweiler (1979), the strain
amplitudes of the ultra-low frequency (ULF) gravity waves generated
by sufficiently massive BHBs may be detected out to cosmological
distances as perturbations in the pulse arrival times of
quiet pulsars (Romani \& Taylor 1983).  With a new generation of
timeable millisecond pulsars (Kaspi, Taylor \& Ryba 1994, hereafter
KTR; Camillo, Foster \& Wolszczan 1994) allowing remarkable
improvements in sensitivity, it seems appropriate to assess the
probability of massive BHB detection.

	In this paper we examine the fate of BHBs in a population of
merger remnants, estimating the frequency of events and using these
results to outline the required pulsar timing sensitivity needed to
effect a detection or constrain the merger hypothesis.  First we
consider inspiral as the average result of repeated interactions with
the surrounding stars (Mikkola \& Valtonen, 1992, hereafter MV; Hills,
1993). The binary loses energy, shrinks and moves faster (hardens),
with moderate increase in eccentricity.  Numerical simulation (Makino
{\it et al.} 1993, hereafter MFOE) seems to confirm this evolution in
general, but warns of the importance of back-action of the BHB on the
surrounding core, neglected in previous analytical investigations.
Since the characteristics of the host core are crucial to the success
of the inspiral, in section 2 we develop a numerical integration of
the evolution of binding energy and angular momentum (hence
eccentricity) during a merger event, which can be applied to a range
of black hole masses and core properties. This approximate trajectory
includes back-action on the core via loss cone depletion (BBR) and
heating.  In Section 2.3, we estimate $f_{SI}$, the fraction of
mergers achieving a successful inspiral due to stellar scattering, by
computing our evolutionary model for a range of core parameters and
BHB properties.  Additional mechanisms that may drive inspiral are
also mentioned in Section 2.3. In section 3 we collect estimates of
the number density of massive black hole cores and galaxy merger rates
to estimate the number of merging sources visible as a function of
gravity wave period and strain amplitude.  In section 4 we compare
these rates to present and anticipated pulsar timing sensitivities and
conclude by commenting on the prospects for further constraining
inspiral events.
\vfill\eject

\section{Simulation of Inspiral}
\subsection{Binary Evolution Formulae}
	It was suggested by Fukushige {\it et al.} (1992b) that the
evolution of a BHB can be computed from the classic Chandrasekhar
formula (Chandrasekhar, 1943) for dynamical friction (DF) on a single
object in a homogeneous background.  Applying an approximate formula
to both holes, they find a large growth in the eccentricity; this was
our motivation to trace energy and angular momentum loss of a BHB.  We
use the results of MV, who track the net effect of an ensemble of
three-body interactions between an equal-mass binary and an intruder
from a background of light objects.  They cast their results in the
form $(\dot a/a^2) = (R_a/v_\infty) \pi G \rho$, where $a$ is the
binary semi-major axis, $\rho$ the density of background stars, and
$v_\infty$ the pre-encounter intruder speed.  They find the numerical
factor $R_a$ to be well fit by $R_a = -6.5/[1+2.44(v_\infty/W)^2]$,
independent of binary eccentricity.  Here $W$ is the r.m.s. relative speed of
the binary elements, {\it i.e.} the relative speed of a circular
binary of equal energy.

	We assume the galactic core to have a Plummer model velocity
distribution, given by
\begin{equation}\label{plumv}
f(v) =  \frac{16}{21\sqrt{3}\pi^2 \sigma^3} \left( 1 - \frac{v^2}{12 \sigma^2}
\right) ^{7/2}.
\end{equation}
Since the one-dimensional dispersion $\sigma$ depends on position in a
 Plummer model, we use the central value $\sigma_0$.  Using the
 fitting function, we calculate (as a function of $\sigma_0$) $\langle
 \sigma_0 R_a/v_\infty \rangle _{\sigma_0}$, the average over Plummer
 distributed $v_\infty$.  This gives
\begin{equation}\label{edot}
\dot E = \frac{G M_1 M_2 \pi \rho}{2 \sigma_0} \left \langle \frac{\sigma_0
R_a}{v_\infty}
\right \rangle _{\sigma_0}
\end{equation}
for the evolution of the binary energy.  In Fig. 1, we compare
(\ref{edot}) to the energy loss rate of a circular binary due to DF,
applying the Chandrasekhar formula (c.f. Binney \& Tremaine, 1987)
independently to each hole.  In the $W<\sigma_0$ limit, the MV formula
agrees by construction with the analytic calculation of Gould (1991);
it thus differs from the DF result in the form of the Coulomb
logarithm $\Lambda$ (see Gould for a discussion).  For the DF
calculations of Fig. 1, we take the maximum impact parameter $b_{max}$
contained in $\Lambda$ to be a function of the binary period, as
discussed in the next section.  In the $W \gtrsim \sigma_0$ regime which
proves most important in our calculation, we see in the figure that DF
does not approximate well the more physically grounded and
asymptotically sound three-body results.

	For the eccentricity evolution, MV give results in the form
\begin{equation}\label{dxda}
\frac{dX}{d\ln a} = X^{n + 1/2} - X \;\;\;\;\;\;\;\;\;\;\;\;
{\rm where}\;\;  X \equiv 1-e^2, \;\;\;\; {\rm and}\;\;
n = \frac{8 \left( \frac{v_\infty}{W}\right) ^{1.5}-0.3 X}{1+16\left(
\frac{v_\infty}{W}\right)^{1.5}}
\end{equation}
For each $X$ and $W$ we again average over $v_\infty$ with
distribution (\ref{plumv}), and convert (\ref{dxda}) to a time
derivative, giving
\begin{equation}\label{xdot}
\dot X = \frac{-G^2 M_1 M_2 \pi \rho}{2 E \sigma_0} \left\langle\frac{\sigma_0
R_a}{v_\infty}
\right\rangle_{\sigma_0} \left\langle X^{n+1/2} -X \right\rangle
_{X,W/\sigma_0}
\end{equation}
In Fig. 1 we also compare the amount of eccentricity growth of the
binary as a function of energy for DF and MV.  Again, DF does not seem
to be a suitable approximation.  Since MV treat only the equal-mass
binary, we impose a correction factor to force agreement with Gould's
analytic mass ratio scaling for $M_1 \neq M_2$.  We use $M_1 \geq M_2$
throughout.

\begin{figure}[htb]
\plotone{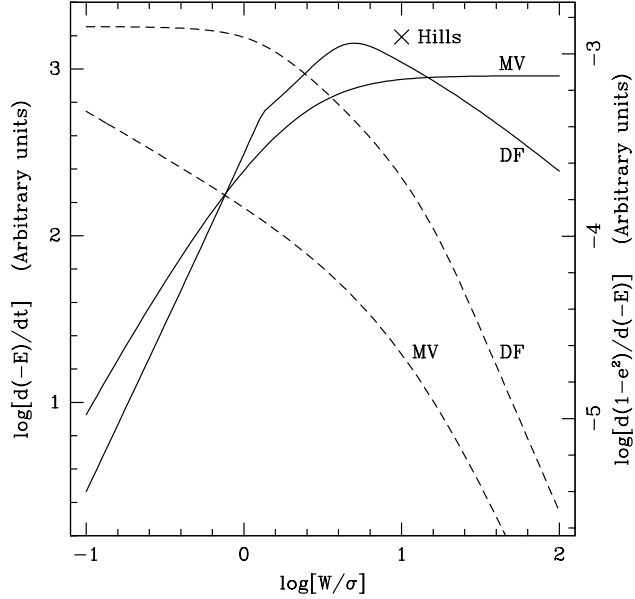}
\caption{Comparison of binary evolution formulae given by dynamical friction
 and by Mikkola and Valtonen (1992).  The left scale applies to the
solid lines giving $\dot E$, and the one value in this range from
Hills (1983); the right scale is for the eccentricity growth rate at
eccentricity of 0.9 (dotted lines).  To calculate $b_{max}$ and
$\Lambda$ for the DF formulae we use $M_1 = M_2 = 1.0 M_9$, $r_c =
200$pc and $\sigma = 300{\rm kms}^{-1}$.  }
\end{figure}

	In following the orbit evolution, we also include the energy and
angular momentum losses to gravitational radiation (GR): (Peters 1964)
\begin{equation}\label{greq1}
\left \langle \frac{dE}{dt} \right \rangle = - \frac{32}{5}
\frac{G^4 M_1^2 M_2^2 (M_1 + M_2)}{c^5 a^5 (1-e^2)^{7/2}}
\left( 1 + \frac{73}{24} e^2 + \frac{37}{96} e^4 \right),
\end{equation}
\begin{equation}\label{greq2}
\left \langle \frac{dL}{dt} \right \rangle = - \frac{32}{5}
\frac{G^{7/2} M_1^2 M_2^2 (M_1 + M_2)^{1/2}}{c^5 a^{7/2} (1-e^2)^2}
\left( 1 + \frac{7}{8} e^2 \right).
\end{equation}
Once these terms dominate, the BHB will circularize.  When the
time scale for binary evolution by GR alone is less than $10^{10}$
years, we call this the GR regime.

\subsection{Properties of the Core}

	We use a Plummer model core of stars with mean mass $m_s$,
core radius $r_c$ and isotropic central velocity dispersion
(\ref{plumv}).  The core stars have total potential
energy W and kinetic energy K, satisfying the virial relation $2K + W
= 0$.  With
\begin{equation}\label{W}
W(N_s,r_c) = -\frac{3\pi}{32} \frac{G (N_s m_s)^2}{r_c} - \frac{G N_s
m_s (M_1+M_2)}{r_c} ,
\end{equation}
we use $K(N_s,\sigma_0)= (9\pi/32) N_s m_s \sigma_0^2$ and the virial
relation to find

\begin{equation}
N_s =  \frac{6\sigma_0^2 r_c}{G m_s} - \frac{32}{3 \pi} \frac{M_1 +
M_2}{m_s}.
\end{equation}
The core mass is $M_c = N_s m_s$.

	Only a subset of the core stars can effectively exchange energy
with the binary.  According to Heggie (1975), these are the stars whose
interaction time with the binary is less than its orbital time scale.  This
criterion defines a maximum impact parameter $b_{max}$ (a function of
the binary period) which we also used in the DF calculation for
Fig. 1. At any distance from the center of the core, the condition on
impact parameter selects stars whose velocities are within a cone aimed
towards or away from the center, the so-called loss cone (Frank
\& Rees 1976).  For our core, the fraction of stars with impact parameter
at most $b_{max}$ is found by direct integration to be
\begin{equation}\label{navail}
\frac{N_{avail}}{N_s} = \frac{b_{max}^2}{b_{max}^2 + r_c^2}.
\end{equation}
Since scattered stars enter a solid angle at a rate proportional to the
solid angle size, stars can enter this loss cone no faster than the
rate $N_{avail}/t_{relax}$.  Thus $t_{relax}$ is also the loss cone
repopulation time,
\begin{equation}\label{lctime}
t_{lc} = t_{relax} = \left( \frac{N_s}{8\ln \left( \frac{r_c\sigma_0^2}{Gm_s}
\right)} \right) \left( \frac{r_c}{\sigma_0} \right) \simeq 3.7\times 10^{13}
 {\rm yr} \left( \frac{N_s}{10^{10}}\right) \left( \frac{r_c}{200{\rm
pc}}\right) \left( \frac{300 {\rm kms}^{-1}}{\sigma_0}\right)
\end{equation}
(Binney \& Tremaine 1987).

	Once encounters with the binary deplete the loss cone (BBR),
the binary evolution rate will be limited by the rate at which stars
enter the loss cone, and the amount of energy each star can take from
the binary.  In the $W > \sigma_0$ limit, MV find the change in binary
energy with each stellar encounter to be $\langle
\delta E \rangle \sim 2 E m_s/(M_1+M_2)$.  This is in good agreement with
the results at several mass ratios of Roos (1988) and Hills (1983).
Writing this as $\langle \delta E \rangle = (M_2/M_1)m_s v_{2c}^2$, where
the circular velocity $v_{2c}$ of the lighter binary component is roughly
the closest approach velocity of a star, we pass to the $W\leq
\sigma_0$ regime by using $\sigma_0$ for $v_{2c}$.  The energy the loss cone
can take away without repopulation is thus $E_{avail} = N_{avail} \langle
\delta E \rangle$, and from (\ref{lctime}), energy enters the loss cone at
a rate $(dE/dt)_{in} = E_{avail}/t_{relax}$.  If each star which
interacts with the binary is scattered into a random orbit, a fraction
$N_{avail}/N_s$ of stars will be returned to the loss cone.  For a
loss cone population in steady state, then, the limiting binary
evolution rate is
\begin{equation}
\left(\frac{dE_{bin}}{dt}\right)_{max} = \frac{N_s}{N_s-N_{avail}}
\left(\frac{dE}{dt}\right)_{in} = \frac{b_{max}^2 + r_c^2}{r_c^2}
\langle \delta E \rangle \frac{N_{avail}}{t_{relax}}
\end{equation}
This rate is generally much smaller than that due to unrestrained
dynamical friction.  We approximate the situation by using the average
core density in (\ref{edot}) and (\ref{xdot}) until $E_{avail}$ drops
below the binding energy of the binary.  From this point on the binary
will affect the loss cone stellar density significantly, and we impose
the steady-state evolution rate.  Invariably this `loss cone
catastrophe' results in a binary evolution time scale
$E_{bin}/\dot{E}_{bin}$ much longer than the Hubble time.  Similar but
less specific treatments have been presented by BBR and Roos (1981).

	We also treat eviction of stars from the core, found to be
significant in the N-body work of Makino {\it et al.} (1993, hereafter
MFOE).  We assume all stars which interact with the binary after
$v_{2c}$ is larger than the Plummer central escape velocity $v_{esc} =
\sqrt{12} \sigma_0$ receive enough recoil velocity to be kicked out
of the core.  We calculate the resultant change in the core energies
assuming rapid revirialization, readjusting $r_c$ and $\sigma_0$.  The
binary also heats the core before becoming hard, but this effect is
small.

	At the beginning of the simulation, the binary's semi-major
axis is the binding radius $r_b$, defined such that the mass of stars
within $r_b$ is less than $M_1$.  Its center of mass is at the core
center, and it has some initial eccentricity $e_i$.  We assume the
core is not rotating after the merger, since N-body merger simulations
show very small angular momentum in the remnant bulge (Barnes 1992),
even when progenitor bulges rotate (Hernquist 1993).  A simulation is
terminated unsuccessfully if $E_{bin}/\dot{E}_{bin}$ exceeds $10^{10}$
years, or successfuly if the GR regime is reached.  To estimate the
parameter space available for successful inspiral, we find $\log
(1-e_i^2)$ for the marginally successful trajectory to one part in a
thousand by using ten bisection steps on the initial value.  Some
sample trajectories are shown in Figure 2.  If we ignore back-action
and loss cone depletion the results are identical to those of MV.

\begin{figure}[htb]
\plotone{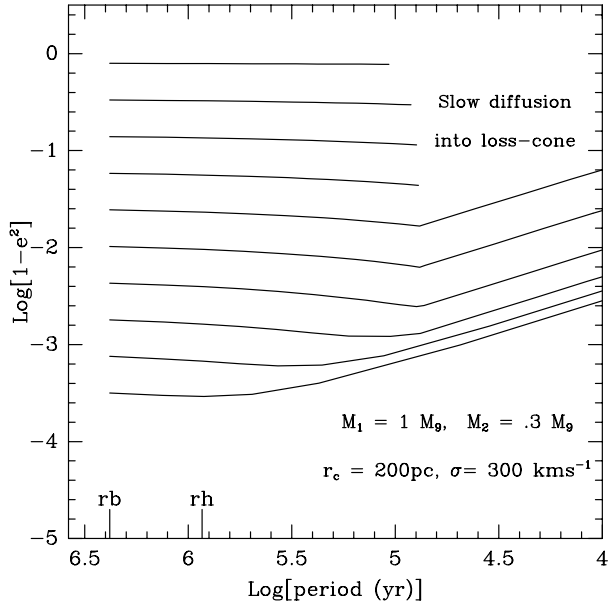}
\caption{Trajectories for two BHB/cluster combinations and several $e_i$.
Slight eccentricity growth and termination at the loss cone limit
are visible, as is recircularization in the GR regime.  Binding radius
$r_b$, and the radius $r_h$ at which $v_{2c}>\sigma_0$ are indicated.}
\end{figure}

	Our simulations confirm the basic behavior described by MFOE.
For a range of initial conditions, interaction with background stars
can cause some increase in the binary eccentricity.  If we choose core
parameters and black hole masses to match those of MFOE, we find loss
cone depletion to occur at binary radii similar to those at which they
terminate their simulation, where they too note a slowdown in binary
evolution which they attribute to a possible loss cone effect.

\subsection{Success Rate and Alternative Merger Mechanisms}

For all pairs of binary member masses drawn from the set $\log
M/M_\odot$ = \{10,9.5,...,7.0,6.5\} we calculated the average initial
critical eccentricity required for successful inspiral in 80 galactic
cores, drawn from a smoothed distribution based on the data
set of Lauer (1985).  We chose the distribution by noting that the
binned data show $N(M/10^{9}M_\odot) = {\rm exp}(-M_9/22)$
(presumably the tail of a Schechter function); matching this and the
$N(r_c)$ of the data set gives our core distribution in the
$r_c-\sigma_0$ plane.  In drawing from the distribution we required that
the stellar mass of the host core be at least twice the total mass of
the holes.

To estimate the fraction of successful inspirals $f_{SI}$ for any
given binary mass, we assume stochastically distributed initial
eccentricities ($P(e) = 2e$), so the inspiral success probability is
$P(e>e_{crit}) = 1-e_{crit}^2$.  This choice of probability may be
somewhat too low since N-body simulations of collisions of
bulge-disc-halo systems (Barnes 1992) show radial final infall of the
bulges; high initial binary eccentricity seems a likely consequence.
Some critical eccentricities for BHB
$(\log\frac{M_1}{M_\odot},\log\frac{M_2}{M_\odot})$ which will later
prove most relevant for pulsar detection are: $ (9.5,9) \rightarrow
.985,\;\; (9,9) \rightarrow .987,\;\; $and $(9,8) \rightarrow .990$.
These yield $f_{SI}$ of 2.9\%, 2.6\% and 2.0\% respectively.  Over the
whole range of binary masses, the $f_{SI}$ are $\sim 1\%-3\% $.  From
the dotted curves in Fig. 1 we see that the dynamical friction
formulae yield much more eccentricity growth, hence earlier arrival in the
GR regime as per equations (\ref{greq1}) and (\ref{greq2}), and larger success
fraction.  We ran calculations with all the above features but using
DF, finding $f_{SI}$ ranging from 1\% to 50\%.  In the case of DF, it
is possible to account for rotation of the stellar core, which was
found to be an important impediment to eccentricity growth in the
simulations of MFOE.  By giving all angular momentum lost by the
binary to the core, and applying the DF formulae to the motion of the
binary relative to the rotating background, we find much less
eccentricity growth, resulting in $f_{SI}$ between 1\% and 16\%.  If
even the modest eccentricity growth in the MV picture, as seen in
Figure 2, is suppressed by spinup of the core, the MV $f_{SI}$ could
be driven slightly lower.

	Clearly stellar action on an isolated BHB in a galaxy core is
of limited efficacy. However, the absence of many double active
galactic nuclei (AGN's) and of orbital variations in the broad line
region, as well as short binary periods inferred from from AGN jet
precession (Roos 1988) suggest that inspiral beyond the loss cone
limit does occur in many cases.  Other means of binary inspiral have
been proposed: gas may be ejected during nuclear activity, taking with
it energy and angular momentum acquired from the binary; or it may be
accreted by the larger hole, causing orbital contraction as $M_1r_2$
remains constant to conserve angular momentum (BBR).

	Roos (1988) has also pointed out that if merger events happen
several times during the lifetime of a galaxy and if, in particular,
mergers are related to the onset of accretion and nuclear activity in
AGNs, then external perturbations due to incoming galactic masses
should accelerate stellar scattering inspiral.  Though a binary which
fails to merge is left evolving on the relaxation (and loss cone
filling) time scale, Roos estimates that by the time a new merging
galaxy is within a few core radii of the remnant pair, the loss cone
can be refilled quickly and repeatedly, driving the original binary to
GR inspiral.  His calculation of the flux of stars entering the loss
cone in this tidal repopulation scheme suggests that the galactic
orbit decays no faster than that of the binary.  If the incoming
galaxy also has a black hole, it seems likely to form a binary with
the post-coalescence hole.

	Accordingly, in the remainder of this paper we follow two
hypotheses: that inspiral occurs with the low probabilities allowed
before loss cone depletion by MV's stellar encounter dissipation, and
that repopulation ensures that inspiral always occurs.  The latter
picture assumes a galaxy formation model involving repeated mergers;
these two cases certainly bracket the actual situation.

\section{Merger Rates and Populations of Nuclear Black Holes}

We wish to compute the rate of mergers observed from some redshift
$z$.  In an Einstein - de Sitter universe ($\Lambda = 0$, $\Omega = 1$)
with the metric ${\rm d}s^2={\rm d}t^2-a(t)^2[{\rm d}\chi^2
+\chi^2{\rm d}\Omega^2]$, let $F_m(z)$ be the number of merger events
at redshift $z$ in the history of today's bright ($L_\ast$) galaxies,
per dimensionless comoving volume per unit redshift.  With this
definition, $\int F_m(z) dz = Nn_{gal}a_0^3$, where $n_{gal}$ is the
current number density of such galaxies and $a_0$ the current scale
factor; N can range from $\sim 1/3$, if only single mergers occur to
form elliptical galaxies, to $\sim 10$ in scenarios where all galaxies
are formed from repeated merging of building blocks.  At redshift $z$
we observe a dimensionless co-moving area $4\pi \chi^2(z)$, with
$\chi(z) =
\frac{2c}{a_0H_0}\left( 1-\frac{1}{\sqrt{1+z}}\right)$.  During a
period $T_z$ at redshift $z$, the comoving volume observed is $V_\chi
= 4\pi\chi^2(c/a) T_z$.  Thus the number of mergers per unit observers'
time is
\begin{equation}
\nu(z) {\rm d}z = \frac{V_\chi F_m(z) {\rm d}z}{T_z(a_0/a)} = 16\pi \left(
\frac{c}{H_0}\right)^2 \frac{c}{a_0^3}
\left( 1-\frac{1}{\sqrt{1+z}}\right)^2 F_m(z) {\rm d}z.
\end{equation}
Note that $a_0$ vanishes in any physical rate because of the
normalization of $F_m(z)$.

	The form of $F_m(z)$ is still quite uncertain.  Burkey {\it et
al.} (1994) find that the population of close galaxy pairs which seem
certain to merge varies as $\sim (1+z)^{3.5\pm.5}$, in the interval
$0 \leq z \leq 0.6$.  Col\'{\i}n {\it et al.} (1994) find that a galaxy density
$\sim (1+z)^{3.8}$ best fits the total galaxy count data in a model
which accounts for photometric evolution.  The 3.5 power law gives
$F_m(z) \sim (1+z)^{2.5}$.  We normalize $F_m(z)$ to the merger
rate implied by the pair counts in Burkey {\it et al.}.  Applying this rate
in the interval $0 \le z \le 1$ implies 40% of all today's bright
galaxies will have suffered a merging event; alternatively if this
rate continues back to the epoch of high quasar activity, $z\sim 3$, then
each bright galaxy will have experienced roughly 5 merger events.  We find:
\begin{equation}\label{freq1}
\nu(z)\,{\rm d}z = 7.6\times 10^{-2} \,{\rm yr}^{-1} h_{50}^{-2}
\left( \frac{n_{gal}}{10^{-3}{\rm Mpc}^{-3}}\right) (1+z)^{2.5}
\left( 1-\frac{1}{\sqrt{1+z}}\right)^2\,{\rm d}z,
\end{equation}
where $H_0 = 50h_{50}{\rm kms}^{-1}/ {\rm Mpc}$.

As an alternative less dominated by early merging, we take merging
rate per comoving volume constant in time, so that $F_m(z) \sim
(1+z)^{-5/2}$. If we assume that merging began at some $z_m$ and
continues to the present, resulting in $N$ mergers per bright galaxy
of number density $n_{gal}$, we find
\begin{equation}\label{freq2}
\nu(z) {\rm d}z = 0.55 \;{\rm yr}^{-1}\left( \frac{N}{h_{50}^2}\right)\left(
\frac{n_{gal}}{10^{-3}{\rm Mpc}^{-3}}\right)
\left( \frac{\frac{3}{2}(1+z_m)^{3/2}}{(1+z_m)^{3/2}-1} \right)
\left( \frac{[1-(1+z)^{-1/2}]^2}{(1+z)^{5/2}} \right) {\rm d}z
\end{equation}

	If accretion onto massive central black holes is the source of
AGN luminosity, the population of remnant holes can be estimated from
models of AGN evolution (Cavaliere and Padovani, 1988; Small and
Blandford, 1992).  Recent HST detections of kinematic evidence for
massive compact objects in nearby galaxy cores support this scenario.
We adopt here Small and Blandford's more conservative model IA, with
a flat local Seyfert luminosity function.  This gives the black hole
number density spectrum
\begin{equation}\label{bhspec}
N_R(M_9){\rm d}M_9 = \left\{   \begin{array}{ll}
7.0 \times 10^{-7} M_9^{-1}{\rm d}M_9 \;\;\; {\rm Mpc^{-3}}  & \;\;\;\;\log M_9
< -1.6 \\
1.2\times 10^{-5} M_9^{-1.4}{\rm d}M_9\;\;\;{\rm Mpc^{-3}} & \;\;\;\;-1.6<\log
M_9<0.45\\
1.1 \times 10^{-4} M_9^{-3.5}{\rm d}M_9 \;\;\;{\rm Mpc^{-3}} & \;\;\;\;0.45 <
\log M_9
\end{array}
\right.
\end{equation}

Comparing the above to a Schecter luminosity function of bright
galaxies $N(L){\rm d}L = 2.5 \times 10^{-3} h_{50}^3 (L/L_\ast)^{-1.1}
e^{-(L/L_\ast)}\, {\rm d}(L/L_\ast) {\rm Mpc}^{-3}$ (de Lapparent {\it
et al.}  1989), we found that if all massive black holes are in
bright galaxies, the integral of equation (\ref{bhspec}) implies that
21\% of $L > L_\ast$ galaxies contain a black hole with $M > 10^{6.5}
M_\odot$ today.  Conservatively, we hold the co-moving number density
of holes fixed: if there are $N_m$ subunits at redshift $z$ that will
become a bright galaxy, then the number of BHs per subunit is
$0.21/N_m$. We note that the high inspiral rates ($\sim 1/3-10 {\rm
yr}^{-1}$) estimated by Fukushige, {\it et al.} (1992a) are based on
excessively optimistic assumption of an $M > 10^8M_\odot$ black hole
in each of $\sim 10$ galaxy subcomponents forming an elliptical. In
their picture there are 10 gravity wave chirps emitted for each bright
elliptical seen today.  As we shall see, such event rates are not
consistent with bounds from pulsar timing.

\section{Gravitational Wave Amplitude of BHB Inspiral}

	For a circular binary of orbital period $P_b$, reduced mass $\mu$
and total mass $M$, Thorne (1987) gives a characteristic strain amplitude
averaged over direction and polarization of
\begin{equation}
h_c = 8 (2/15)^{1/2} \mu (2 \pi M/P_b)^{2/3}/r
\end{equation}
for the emitted gravitational wave from a circular ($e=0$) binary.
In units convenient to the BHB problem this is
\begin{equation}
h_c = 1.8 \times 10^{-15} M_1 M_2 (M_1+M_2)^{-1/3} r_{Gpc}^{-1} [g^{1/2}
(e,n)/n] P_w^{-2/3}
\end{equation}
where masses are in units of $10^9M\odot$, distances are in
kiloparsecs, $g(e,n)$ gives the fraction of the wave power in the
$n^{\rm th}$ harmonic as a combination of Bessel functions (Peters \&
Mathews 1963) and the observed gravitational wave period in years is
$P_w= P_b(1+z)/n$.  Only the $n=2$ harmonic is non-zero for a circular
orbit.

	Since the gravity wave flux scales as $h^2\omega^{-2}$,
falling off with the luminosity distance as $d_L^{-2}$, one can write
the gravity wave amplitude from a source at $z$ emitting waves that
have period $P_{w}$ at redshift 0 in an Einstein-deSitter universe:
\begin{equation}\label{hamp}
h_{-15}(z) = 0.29 M_1 M_2 (M_1+M_2)^{-1/3} h_{50} [g^{1/2}(e,n)/n] P_w^{-2/3}
{{(1+z)^{2/3}} \over {[1-(1+z)^{-1/2}]}}
\end{equation}
in dimensionless units of $10^{-15}$. The characteristic lifetime for
the source at this period is
\begin{equation}
\tau_{GR}=1.3 \times 10^4 M_1^{-1}M_2^{-1}(M_1+M_2)^{1/3}f(e)^{-1}
[n\,P_w/2(1+z)]^{8/3} \,{\rm y}
\end{equation}
where $f(e)=(1+\frac{73}{24}\,e^2 + \frac{37}{96}\,e^4)/(1-e^2)^{7/2}$,
as in Eq. (\ref{greq1}).

	We wish to estimate, for a set of assumptions about the
population, the expected number of BHBs detectable at a given strain
sensitivity $h_{-15}$. Inverting equation (\ref{hamp}) gives
$z(h;M_1,M_2,P_w,e)$.  We can combine this with the merger frequency
rate, two independent draws from the black hole mass function
(\ref{bhspec}), the successful inspiral fraction for these hole masses
(averaged over cores that might contain such holes), and the lifetime of the
resulting binary at the indicated orbital period to get the number of
visible gravity wave sources in a given amplitude range:
\begin{equation}
{{{\rm d}N(h,M_1,M_2,P_w,e)}\over {{\rm d}M_1{\rm d}M_2}} =
\nu(z){\rm d}z{{N_R(M_1)N_R(M_2)} \over {N_{L_\ast}^2}}
\langle f_{SI}(M_1,M_2)\rangle \tau_{GR}(M_1,M_2,P_w,e,z)
{{{\rm d}h} \over {{\rm d}z}}.
\end{equation}
Integration of this equation over $M_1$, $M_2$ and a range of $h$
gives the source population estimates shown in Figure 3a for waves of
period $P_w = 10$y. In practice, we find that even for inspiral driven
by stellar encounters, the characteristic eccentricity of the binary
decreases to $\la 0.3$ by the time it reaches the range observable
with pulsar timing, so restriction to circular orbits is reasonable
for estimating fluxes. Depending on assumptions, we see that gravity
waves should be detectable once a strain sensitivity of $\sim
10^{-15}$ is reached. The properties of the detected BHBs can also be
estimated by weighted sums over this population; for the case of
$f_{SI}=1$, and mergers proceeding as in equation (\ref{freq1}),
average characteristics of the detected binaries are shown in Figure
3b.  It is also worth noting that with these parameters, we expect
$\sim 1$ chirp per $\; 10^4$yr with an amplitude greater than
$10^{-15}$ at periods $P_w \sim 10^4$s, substantially lower than the
rates quoted by Fukushige {\it et al.} (1992a).

\begin{figure}[htb]
\plottwo{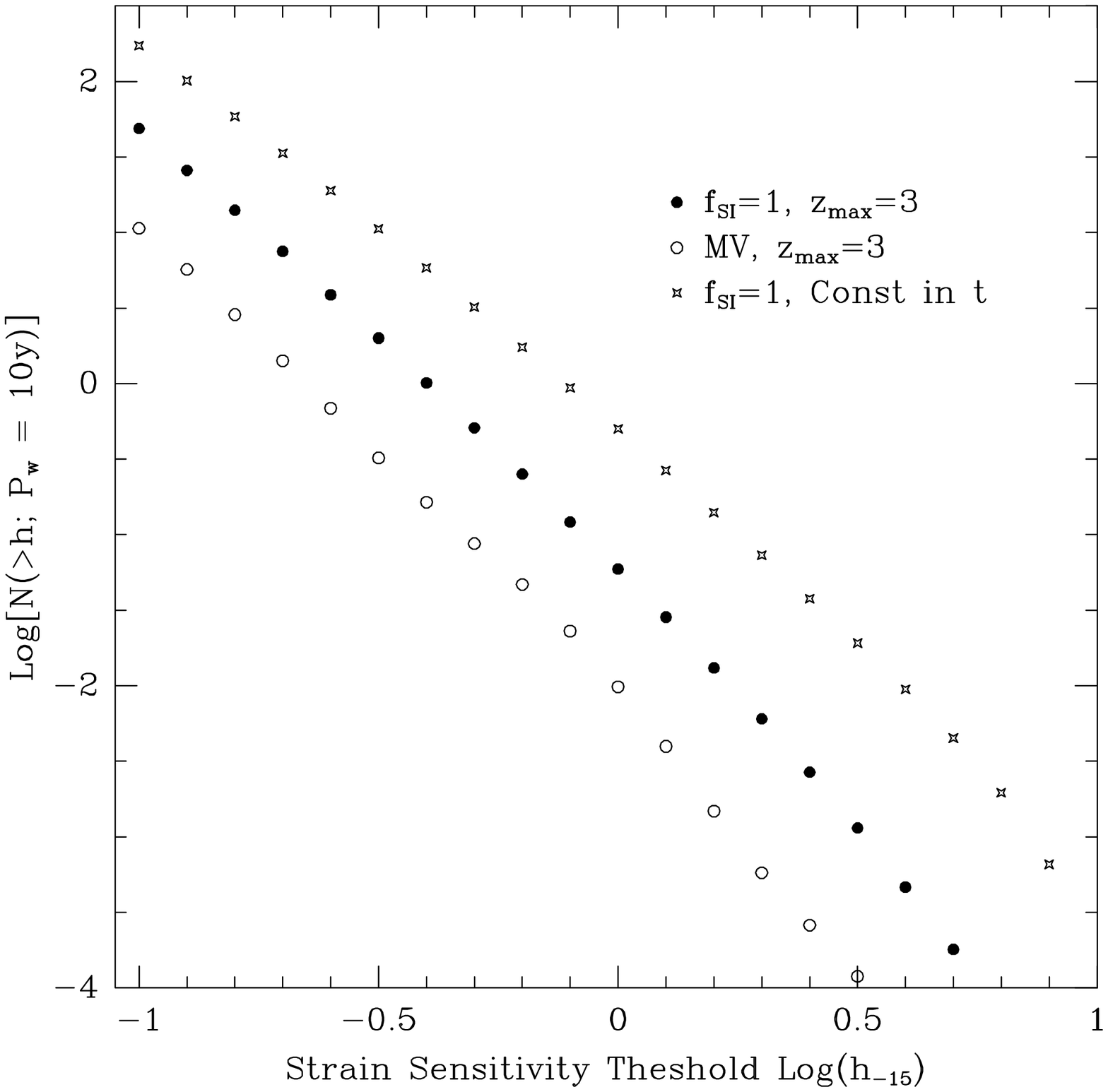}{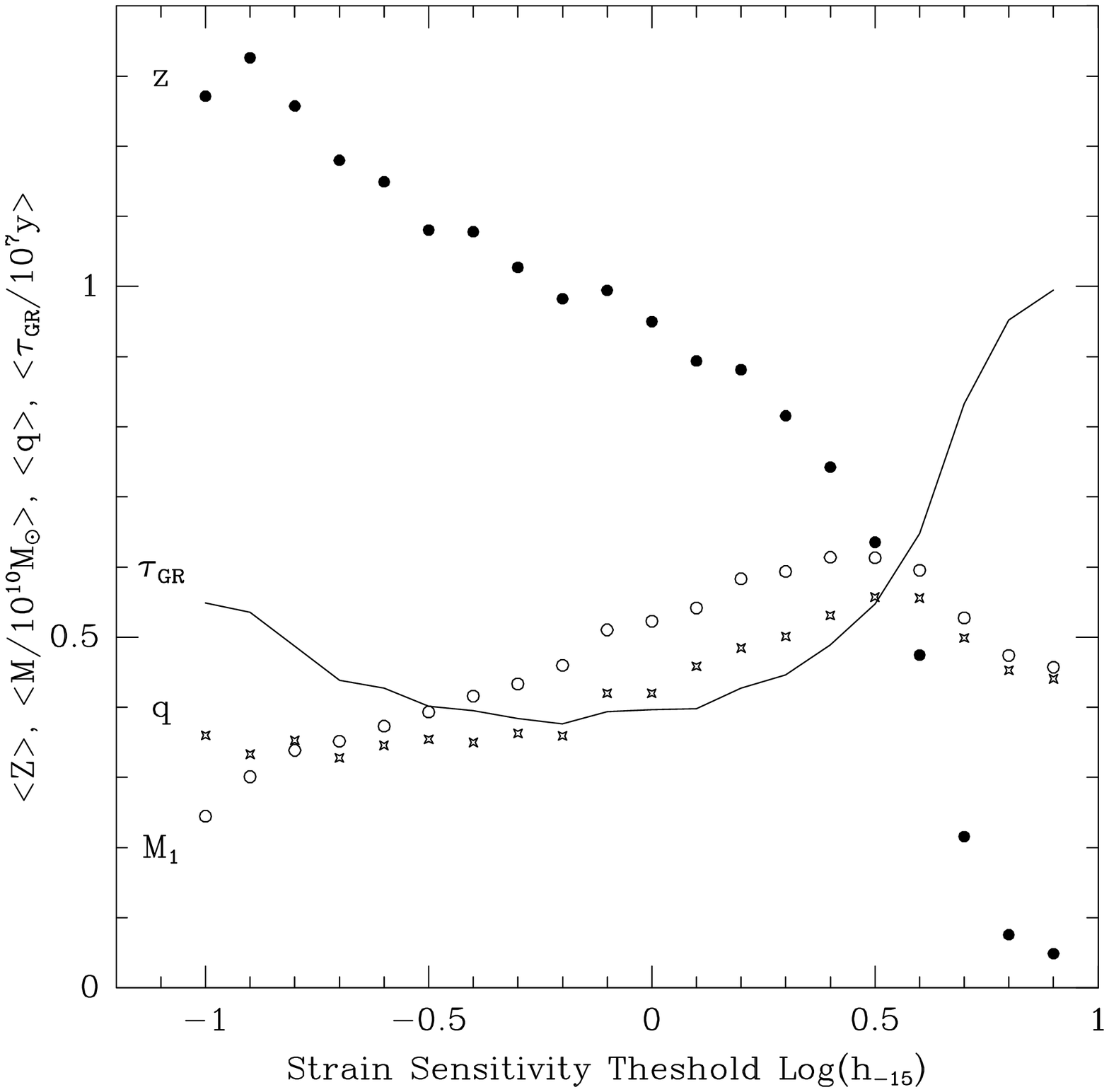}
\caption{Left -- Source intensity distributions for several merger models.
The filled dots give the distribution for tidally repopulated merging
and open circles show numbers for evolution slowed by loss cone
depletion, with the merger rate (13).  Crosses show the population
from mergers constant in $t$.  Right -- weighted observed source
properties for $f_{SI}=1$ mergers following rate (13).  The
binary mass ratio is q, and the GR inspiral time is $\tau$.   }
\end{figure}

\section{Conclusions and Observational Prospects}

	Millisecond pulsar timing has been used to place strong bounds
on the energy density in a stochastic background of ultra-low
frequency gravitational waves (KTR and references therein).  For
example, these studies have placed an energy density limit $\Omega_g <
2 \times 10^{-7} h_{50}^{-2}$ and have helped to rule out various
exotic cosmologies, such as structure formation seeded by cosmic
string loops. With a number of high-quality pulsars now being timed,
it is worth considering whether astrophysical sources, such as BHBs,
are within reach.  In Figure 4 we show a simulated gravity wave
spectrum computed from the amplitude number distributions (as in
Figure 3a) for a range of wave periods. Two models are shown, the
first with inspiral mediated by tidal repopulation ($f_{SI}=1$), the
second for evolution limited by loss cone depletion.  Amplitude
distributions were computed at each frequency; then the power was
integrated over the low $h$ sources and Monte Carlo sampled from the
rare bright sources. The corresponding amplitude is given in
Figure 4, which thus shows the expected BHB ultra-low frequency
gravity wave background.

\begin{figure}[htb]
\begin{center}
{
\leavevmode
\epsfxsize=10cm
\epsfbox{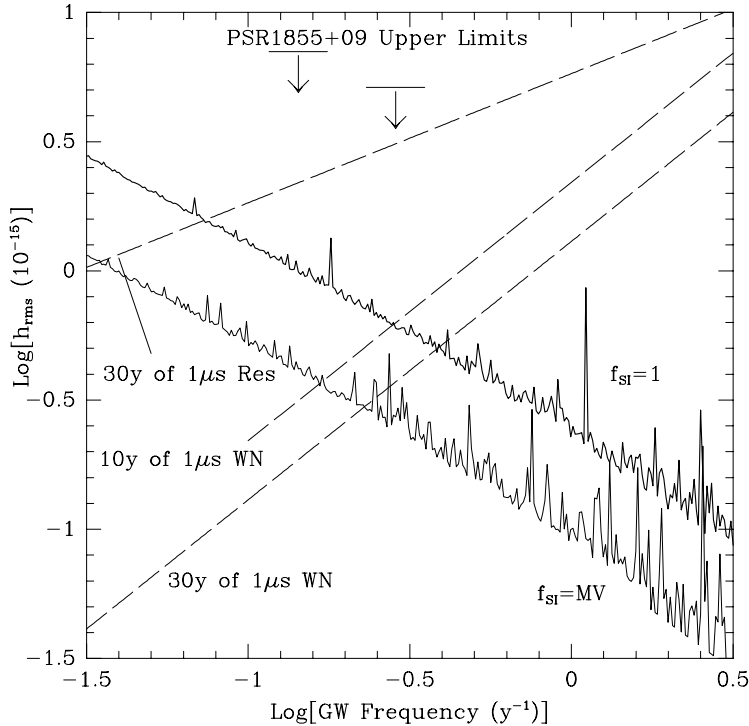}
}
\caption{ Simulated ULF gravity wave backgrounds from $f_{SI}=1$ mergers
and loss cone limited stellar encounter (MV) mergers. Bright nearby
sources rise above the background. Approximate sensitivities from
anticipated pulsar timing experiments are shown as dashed lines (see
text).}
\end{center}
\end{figure}

	If one makes $N_{obs}\sim 20$ arrival time measurements per year over
a period $T_{obs}\sim 10{\rm y}$ with an accuracy of $\delta t \sim 1
\delta_{-6} \mu$s,
then one can place a limit on the strain amplitude of a passing gravity wave
with period $P_w$ of roughly
\begin{equation}
h \sim {{\delta t} \over P_w} (N_{obs}T_{obs})^{-1/2} \sim
2 \times 10^{-15} \delta_{-6} (N_{20}T_{10})^{-1/2} P_y^{-1}.
\end{equation}
For this estimate to hold we must have $T_{obs} \ga P_w$, so that
fitting pulsar parameters does not significantly absorb any gravity
wave signal (Blandford {\it et al.} 1984).  This also assumes that
the pulsar timing residuals are distributed as white noise. According
to this estimate, intensive long-term timing programs can reach the
sensitivity needed to detect the brightest BHB gravity wave sources at
periods near 10y.

	Present timing results (KTR) show that PSR1937+21 which has
been monitored for over eight years shows significant unmodeled timing
noise, presumably due to rotational variations intrinsic to the
pulsar.  PSR1855+09, on the other hand, shows random variations at its
$\sim 0.8\mu$s arrival time accuracy with over seven years of timing;
these variations are consistent with perturbations arising from
instability of the best terrestrial clocks. PSR1855+09 residuals are
constraining gravity wave sources at an amplitude of $h \approx 5
\times 10^{-15}$ (Fig. 4).  Thus with the best present data, we
estimate a chance $\sim 0.001$ of detecting a merging BHB. However,
with a factor of $\sim 5$ increase in sensitivity (and a slightly
increased experiment duration) we can anticipate detecting the
brightest BHB gravity wave sources, for reasonable population
assumptions.  Whether this sensitivity increase can be effected is
uncertain.  In the last few years, timing programs have been initiated
on several new pulsars ($e.g.$ PSR J1713+0747, PSR J0437--4715) that
provide sub-$\mu$s timing residuals. However, to take advantage of
this precision it will be necessary to obtain improved atomic clock
standards or time one pulsar against another. Finally, to reach
interesting detection sensitivities we require the arrival time
residuals to integrate down as white noise over the $\sim 10$y
periods. Clearly, intrinsic instabilities in PSR1937+21 prevent this;
fortunately timing noise appears to correlate with period derivative
and several of the new pulsars have period derivatives 10 - 20 times
smaller than that of PSR 1937+21. It should be noted that the required
single order of magnitude improvement represents a much better
prospect than most other gravitational radiation search techniques!

       In summary, we have developed a model of the inspiral of two
massive black holes in a galaxy core driven by stellar encounters. We
recover some of the behavior described by earlier workers, though the
large eccentricity needed to circumvent the loss cone catastrophe in
our calculations must be present {\it ab initio}.  Our simplified sum
allows computation of this process in a range of cluster cores, and we
see that simple stellar encounter dissipation is only effective in
1\%-3\% of all BHB-producing merger events. Nonetheless, it seems
likely that other processes will ensure that inspiral occurs in the
majority of cases.  Turning to the rate of galaxy mergers, and the
fraction of merging cores containing high mass black holes, we
estimate the event frequency and typical gravity wave amplitude
expected from a cosmological population of merging cores with central
BHBs. Simulating the gravity wave spectrum produced by this population
shows that, while no detection of ULF GR sources is expected to date,
moderate improvements in present sensitivities will make detection of
waves from binary sources with periods of $\sim10$yr possible {\it
via} timing observations of millisecond pulsars. Detection of
correlated gravity wave signals in the arrival times of several
pulsars would, of course, constitute an exciting confirmation of this
astrophysical class of gravity wave sources. However, even upper
limits a factor $\sim 5$ lower than present bounds can constrain the
population and merging behavior of massive BHBs throughout the
universe.

\bigskip
\bigskip
	It is a pleasure to acknowledge useful discussions with Andrew
Gould and Douglas Richstone, and helpful comments by the referee.  RWR
was supported in part by an Alfred P.  Sloan fellowship, and MR in
part by a fellowship from the National Science and Engineering
Research Council of Canada.

\vfill\eject

\end{document}